\documentclass[preprint]{emulateapj}
\journalinfo{Submitted for publication in ApJ}
\submitted{Submitted for publication in ApJ, \today}
\usepackage{graphicx}
\usepackage{fancyref}
\usepackage[pdftex,plainpages=false]{hyperref}
\usepackage{amsmath}
\shorttitle{Effects of Angular Momentum on Halo Profiles}
\shortauthors{Lentz, Quinn, Rosenberg}

\begin{document}

\title{ Effects of Angular Momentum on Halo Profiles }

\author{Erik W Lentz\altaffilmark{1}}
\author{Thomas R Quinn \altaffilmark{2}}
\author{Leslie J Rosenberg \altaffilmark{1}}

\altaffiltext{1}{Physics Department, University of Washington,
                 Seattle, WA 98195-1580;       
		 {\tt lentze@phys.washington.edu, ljrosenberg@phys.washington.edu}}
\altaffiltext{2}{Astronomy Department, University of Washington, 
                 Seattle, WA 98195-1580;       
		  {\tt trq@astro.washington.edu}}

\begin{abstract}
The near universality of DM halo density profiles provided by N-body simulations has proven to be robust against changes in total mass density, power spectrum, and some forms of initial velocity dispersion. In this letter we study the effects of coherently spinning up an isolated DM-only progenitor on halo structure. Halos with spins within several standard deviations of the simulated mean ($\lambda \lesssim 0.20$) produce profiles with negligible deviations from the universal form. Only when the spin becomes quite large ($\lambda \gtrsim 0.20$) do departures become evident. The angular momentum distribution also exhibits a near universal form, which is also independent of halo spin up to $\lambda \lesssim 0.20$. A correlation between these epidemic profiles and the presence of a strong bar in the virialized halo is also observed. These bar structures bear resemblance to the radial orbit instability in the rotationless limit.

\end{abstract}


\section{Introduction}
\label{Introduction}

The $\Lambda$CDM model is considered by many to be the most successful to date in predicting cosmological structure formation in our universe, and is often referred to as the standard model of cosmology (see \citet{Planck2014,Planck2015} for a review). $\Lambda$CDM matches large scale structure observations down to a few Mpc \citep{Eisenstein2005}. Even below a Mpc, $\Lambda$CDM provides vast insight into the formation of clusters, galactic halos, and sub-halos; but its limitations also become more evident.

One of these limitations is the so-called core-cusp problem, which can be described as an inconsistency between the inner density profiles of halos found in numerical simulation and those extracted from galactic observations \citep{Moore1994, Flores1994}, with the density inside a scale radius being much shallower than expected \citep{Walker2010}. Approximating the density $\rho$ scaling as a power law at small radii ($\rho \propto r^{\alpha}$ for $r \to 0$), Cold Dark Matter (CDM) numerical predictions show a nearly universal profile with $\alpha = -1$ or steeper, which closely resembles the curve put forth by \citet{Navarro1996a} \citep{Navarro1996b, Moore1994, Moore1998, Moore1999}. Current observations, however, provide scaling much closer to $\alpha = -0.5$ or even flat ($\alpha = 0.0$) \citep{Navarro2010}. Further, the central density of a dwarf DM halo is observed to be constant ($\sim 10^8 M_{\odot} \text{kpc}^{-3}$) over multiple orders of magnitude in halo mass \citep{Strigari2008,Gilmore2007,Hayashi2015}.

In response to the core-cusp problem and other concerns, an abundance of new physics has been proposed to resolve a new model with the data. These theories' effects range from heating the dark matter by power truncation \citep{Governato2015, Moore1999}, giving the DM the ability to repulsively self-interact through means beyond gravity \citep{Moore2000,SpergelSteinhardt2000,Burkert2000,Dave2001,Fry2015a,Fry2015,Vogelsberger,Zavala}, to having DM structure include its own sector of forces and other physics \citep{SUSY, Axions}. These new physics create a model landscape, much of which is currently unexplored.

Attempts have been made to understand the nature of the universal profile, which we will approximate using the NFW profile \citep{Navarro1996a, Navarro1996b}. Statistical techniques applied to this understanding include analysis of approximate integrals of motion during entropy maximization \citep{Pontzen2012}, semi-analytic equilibrium phase-space density estimation \citep{Taylor2001, Barnes2006}, {and mass accretion history \citep{Ludlow2013}}. Under various constraints, the integral and equilibrium studies produce NFW-like profiles, though there are still gaps in understanding of each technique's motivations, keeping the physical cause of NFW ambiguous. {Accretion history studies have also made progress in identifying the limits of NFW, but still fall short in identifying the fundamental mechanisms responsible for the observed similarity}. Other semi-analytical approaches include spherical shell collapse models, which produce results that appear to challenge universality \citep{Zukin2010a, Zukin2010b}.

The evolution equations governing structure formation are in general non-linear and, as a result, both direct and semi-analytic approaches are used to solve them. To date, semi-analytic methods have probed structure formation before the time of first-crossing, contributing to topics from angular momentum distribution to the cosmic web \citep{Binney1990, Quinn1992, Ryden1987, Casuso2015, Pichon2014, Codis2015}. While these methods cannot probe the current epoch, their form and computational efficiency provide agility when surveying classes of models. To reach virialization, a full non-linear approach is needed to retain any sort of accuracy. Thankfully, powerful numerical schemes have been developed which leverage the ever-increasing computational power available. These techniques include both particle, mesh, and hybrid codes, many with sophisticated optimizations that allow them to scale quite efficiently \citep{ChaNGa1, gadget, hydra, enzo}.

This letter concentrates on how computational methods have been used to study the feature of angular momentum and its role in structure formation \citep{Bellovary, Bullock, vandenBosch2002,Barnes1987,Gardner2000,Herpich2015}. Most angular momentum studies quantify a structure's spin by the parameter $\lambda$ \citet{Peebles1969}
\begin{equation}
\lambda = \frac{J |E|^{1/2}}{G M^{5/2}}
\end{equation}
where $J$ is the magnitude of the total angular momentum, $E$ is the total energy, and $M$ is the total mass of the halo.

Angular momentum analysis of cosmological structures has a decades-long history, with analyses that span over both the linear and non-linear approaches. Chief among their findings is the log-normal distribution to total halo spins, with semi-analytics producing a mean value of $\lambda \sim 0.09$ \citep{Ryden1988} and non-linear means giving $\lambda \sim 0.035$ \citep{Bullock, Barnes1987}, both essentially contained well within the small rotational energy limit.

This letter focuses on how halo formation is affected over the range of possible spin values and the subsequent difference of the density profiles from the universal form. In Section~\ref{Methods}, we describe the series of N-body isolated collapse simulations used to consistently study spin's contribution to a virialized halo. Section~\ref{Results} displays the resulting structures and how they relate to the universal profile. Section~\ref{Discussion} discusses what effects, if any, can be attributed to the angular momentum content of the halo or its progenitor. We end with a summary of these findings and potential directions for future study in Section~\ref{Summary}.

\section{Methods}
\label{Methods}

To isolate the effects of angular momentum on the structure of halos, isolated spherical perturbations with a range of spin values are evolved in an expanding universe to form halos of a prescribed virialized size {at zero redshift}.

Initial profiles of the collapse are generated by the Isolated Collapse Initial Conditions Generator (ICInG) package \citep{ICInG}, made specifically for isolated spherical collapse. The initial condition distributions follow from \citet{Evrard1988} and assume an Einstein-deSitter cosmology. ICInG uses a glass distribution to avoid numerical artifacts and includes the ability to impart angular momentum as well as random motions.

The generator begins with a density profile at a prescribed initial redshift $z_{i}$ given by a top-hat over-density of the form
\begin{equation}
\delta(r) = \frac{\delta_0}{2} \left( 1+ \cos\left(r \pi/R_{s} \right) \right) \label{delta}.
\end{equation}
where $R_{s}$ is the radius of the sphere and $\delta_0$ parameterizes the magnitude of the over-density, both of which help specify the mass of the resultant halo. 

Given a collection of particles initially distributed in the homogeneous sphere, the radii are shifted to match the perturbed density profile
\begin{equation}
r_0 \to r_1 = r_0 \left(1- \delta(r_0)/3\right).
\end{equation}
The velocity distribution is comprised of the radial expansion rate of the embedding universe, a peculiar velocity given by linear growing mode theory \citep{Evrard1988}, a rigid rotor condition parameterized by the spin $\lambda$, and a random velocity seed of magnitude equal to the peculiar velocity added to prevent singular collapse 
\begin{align}
\mathbf{v} &= \mathbf{v}_{\text{hub}} + \mathbf{v}_{\text{pec}} + \mathbf{v}_{\text{rot}} + \mathbf{v}_{\text{rand}} \nonumber \\
&=  H \mathbf{r}_1  - \frac{2}{3}H \delta(r_0) \mathbf{r}_0  + \boldsymbol\omega \times \mathbf{r}_1  + \mathbf{v}_{\text{rand}} ,
\end{align}
where
\begin{equation}
\boldsymbol\omega = \hat{\mathbf{z}} \frac{ 5 \lambda G M^{3/2}}{2 |E|^{1/2} R_{s}^2}. \label{spin}
\end{equation}

The N-body code ChaNGa is then used to evolve the distribution using collisionless dynamics. The code uses a Barnes-Hut tree to calculate gravity, with hexadecapole expansion of nodes. Time-stepping is done with a leapfrog integrator with individual time-steps for each particle. The code base has been thoroughly tested and has contributed to many astronomical topics including DM candidate testing \citep{Kim2014, Governato2015, ChaNGa3}
\footnote{We used the public distribution of ChaNGa available via the UW N-Body Shop GitHub page (https://github.com/N-BodyShop/changa). Literature on its operation can be found on the wiki page (https://github.com/N-BodyShop/changa/wiki/ChaNGa).}.

Although the collapse is scale-free, we have chosen a physical scale to simulate the equivalent of a large MW-sized halo ($M_{\text{vir}} \approx 1.5\times10^{12} M_{\odot}$) starting at $z_i=6$ with an effective initial over-density $\delta_0\approx0.71$ {for all halos save $\lambda=0.00$, where $\delta_0\approx0.84$}. The progenitor particle number is $\sim$million with softening length $\sim 0.4$ kpc, which is more than sufficient to resolve core depletion at the $\sim 4$ kpc scale. The system is evolved for $\sim 10$ Gyr with a force accuracy/node opening criterion $\theta=0.7$ and a time-stepping accuracy such that the time-step $\Delta t < \eta \sqrt{\epsilon/a}$, where $\epsilon$ is the gravitational softening, $a$ is the acceleration of a particle, and $\eta$ is an accuracy criterion; we used $\eta =  0.00013$. Also note that as these are DM-only simulations, the configuration may be rescaled to a halo size of our choosing because Newtonian gravity is scale-free. Cosmological expansion is turned on, but to model the system as isolated, boundary conditions are not periodic.

To survey the effects of angular momentum, we explore the $\lambda$ landscape {at the points $\lambda \in \{0.00,0.03,0.06,0.08,0.10,0.15,0.25,0.50\}$}, where $0.03$ is a typical value for galactic halos \citep{Barnes1987, Bullock, Ryden1988}, and $0.50$ is considered quite unusual. In the next section, we discuss the outcome of these simulations.

\section{Results}
\label{Results}

\begin{figure}[]
\begin{center}
\includegraphics[width=8.5cm]{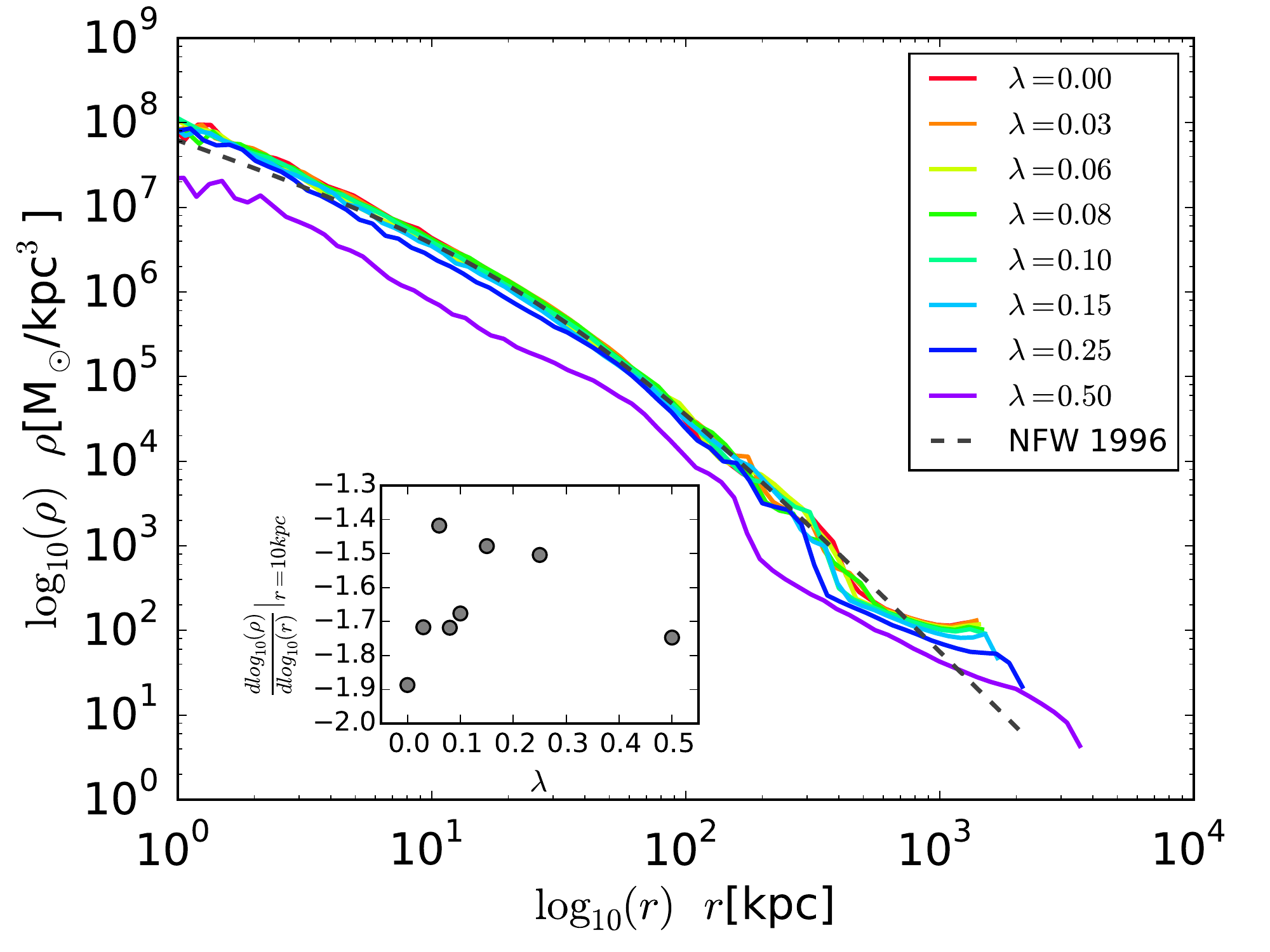}
\caption{Spherically averaged density profile for the various spin values after collapse. The lower spin progenitors form virialized states that match the nearly universal NFW form, save for the notable increase in power law within the scale radius. The reference NFW profile has scale radius $30\text{ kpc}$ and scale density $2.2 \times 10^6 M_{\odot}/\text{kpc}^3$, which were chosen by eye to match the low spin profiles. {The inset plot displays the power law behavior of each profile at $10$kpc.}}
\label{fig:fig1}
\end{center}
\end{figure}

After the progenitor is {evolved to $z \approx 0$, properties such as the radial density profile are calculated}, shown for our halos in Fig.~\ref{fig:fig1}. For the lower $\lambda$ in our range {$\{0.00,0.03, 0.06, 0.08\}$}, the profiles tightly match a universal form as expected. By that we mean they closely follow the NFW \citep{Navarro1996a} density curve with the slightly faster than $r^{-1}$ growth towards the center that has become expected by some for virialized halos \citep{Fry2015, Moore1994, Moore1998, Moore1999}. Surprisingly, the form persists into the higher spin halos until  $\lambda=0.25$, which lies in the $99.9$ percentile of halos according to some N-body studies \citep{Barnes1987, Bullock}. Deviations in the higher spin profiles, $\lambda \in \{0.25,0.50\}$, include {a divergence from the two power law form, a shrinking virial radius, and a lowered density, the last two of which may be partially attributed to the limitations of ICInG in the high spin regime to set the final virial mass. Our definition of the virial radius and mass are in line with \citet{Bullock}, with the virial radius taken to be the distance at which the average interior density equals $200$ times the critical density. The particles within the virial radius at this time are called the virialized component, while the remainder are the non-virialized component.} The relation between the $\lambda$ values supplied to ICInG and the measured spins of the virialized halos are provided in Table~\ref{tab:table1}.

\begin{table}[]
\centering
\title{Halo Spin Values}
\begin{tabular}{ll}
\hline
\multicolumn{1}{|l|}{$\lambda$ (parameter)} & \multicolumn{1}{l|}{$\lambda$ (measured)}  \\ \hline
\multicolumn{1}{|l|}{0.00} & \multicolumn{1}{l|}{0.000}  \\ \hline
\multicolumn{1}{|l|}{0.03} & \multicolumn{1}{l|}{0.035}  \\ \hline
\multicolumn{1}{|l|}{0.06} & \multicolumn{1}{l|}{0.064}  \\ \hline
\multicolumn{1}{|l|}{0.08} & \multicolumn{1}{l|}{0.094}  \\ \hline
\multicolumn{1}{|l|}{0.10} & \multicolumn{1}{l|}{0.109}  \\ \hline
\multicolumn{1}{|l|}{0.15} & \multicolumn{1}{l|}{0.168}  \\ \hline
\multicolumn{1}{|l|}{0.25} & \multicolumn{1}{l|}{0.248}  \\ \hline
\multicolumn{1}{|l|}{0.50} & \multicolumn{1}{l|}{0.381}  \\ \hline
\end{tabular}
\caption{Relation between the $\lambda$ parameter values supplied to ICInG and those measured in the virialized halo.}
\label{tab:table1}
\end{table}

\begin{figure}[]
\begin{center}
\includegraphics[width=8.5cm]{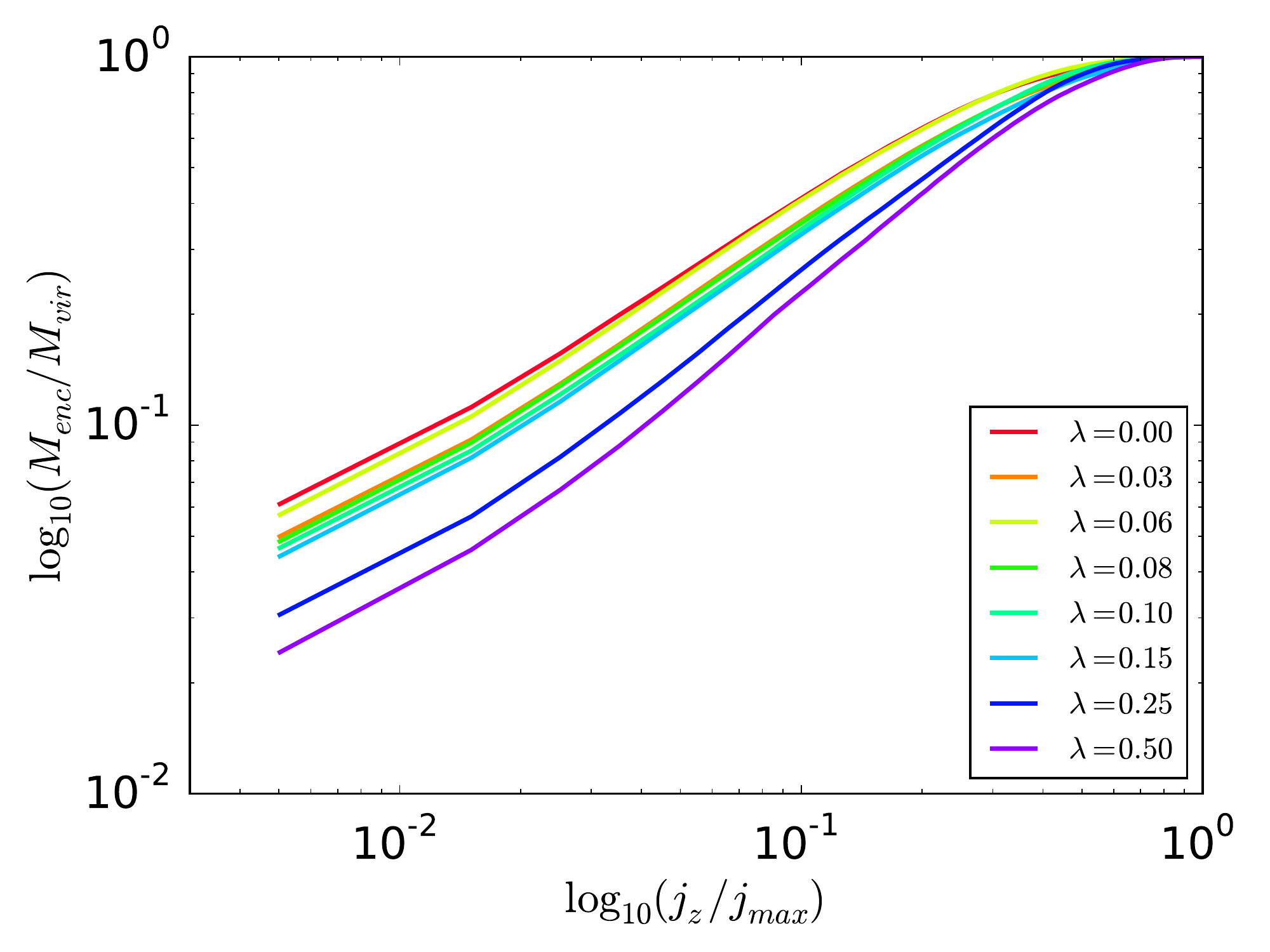}
\caption{Enclosed-mass profile of prograde particles versus the vertical component of specific angular momentum, reminiscent of Figure 4 of \citet{Bullock}, showing the lower spin halos take a similar form while the higher spin halos have a steeper inner slope. A difference in technique may be responsible for the discrepancy: Bullock removed large amounts of retrograde ($j_z<0$) material from the halos before profiling, whereas we have removed every retrograde particle.}
\label{fig:fig2}
\end{center}
\end{figure}

Cumulative mass profiles of prograde particles within the virial radius versus specific angular momentum also show a consistency over the lower spin halos, Fig.~\ref{fig:fig2}, despite their initial differences in rotation rate. \citet{Bullock} (BDK) performed a similar analysis and proposed a universal profile of the form
\begin{equation}
M_{\text{enc}}(j) = M_{\text{vir}} \frac{ j }{j - j_0} \label{Bul_prof}, 
\end{equation}
where $j$ is the z-component of the specific angular momentum as defined by the imposed spin axis \eqref{spin}. This cumulative mass uses only the prograde bodies in the virialized halo, summing the masses of particles with specific angular momentum between $0$ and $j$. {For reasonable $\lambda$, the inner power law ($M_{\text{enc}} \propto j^{\alpha}$ for $j \approx 0$) of the calculated profiles is noticeably shallower with $\alpha \approx 0.7$}. The primary difference between our analysis and BDK's is the presence of some retrograde ($j_z<0$) constituents due to the differences of their binning process. Due to the consolidated profiles over reasonable $\lambda$, we suggest a similar profile shape for the slower halos
\begin{equation}
M_{\text{enc}}(j) = M_{\text{vir}} \left( \frac{j}{j - j_0} \right)^{\alpha} \text{ , } \alpha \approx 0.7 \label{j_prof}.
\end{equation}

\begin{figure}[]
\begin{center}
\includegraphics[width=8.5cm]{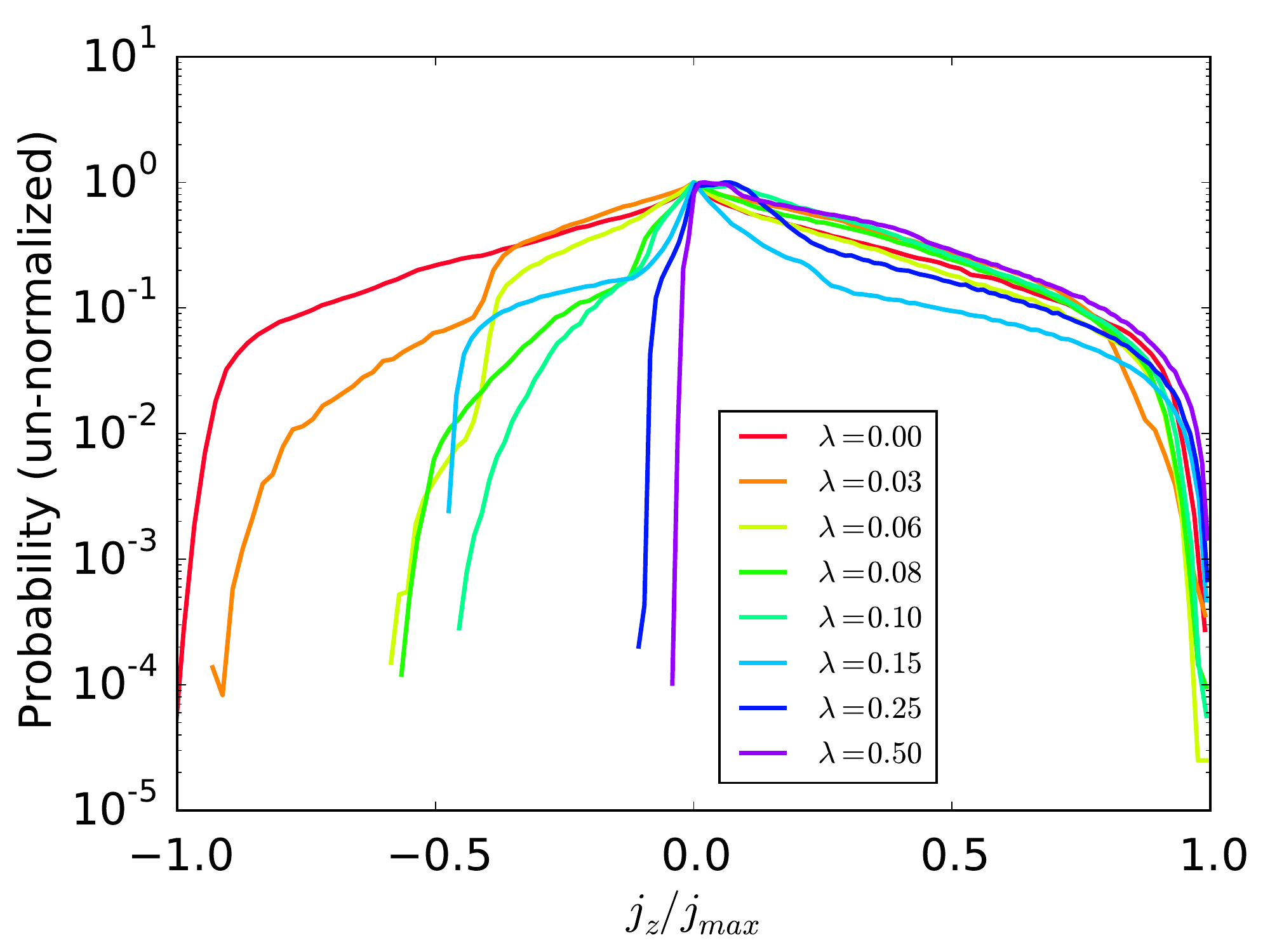}
\caption{Probability density function of the isolated distributions after collapse. The initial configurations for each distribution were well represented by a rotating solid sphere, which would give a distribution with support in positive $j$. The retrograde component ($j_z<0$) to the final distributions implies a breaking of the initial axial symmetry.}
\label{fig:fig3}
\end{center}
\end{figure}

\begin{figure*}[t]
\begin{center}
\includegraphics[width=18cm]{IsolatedCollapse_final_overdensity_snapshots.pdf}
\caption{Integrated face-on over-density views of all collapsed configurations, obtained via azimuthal inverse Fourier transform over cylindrical bins and normalized bin-wise. The radial bar-like structure forming in the slower halos is visible as the over-dense cones, whereas the faster halos' bar's appear increasingly disrupted.}
\label{fig:fig4}
\end{center}
\end{figure*}

\begin{figure}[]
\begin{center}
\includegraphics[width=8.5cm]{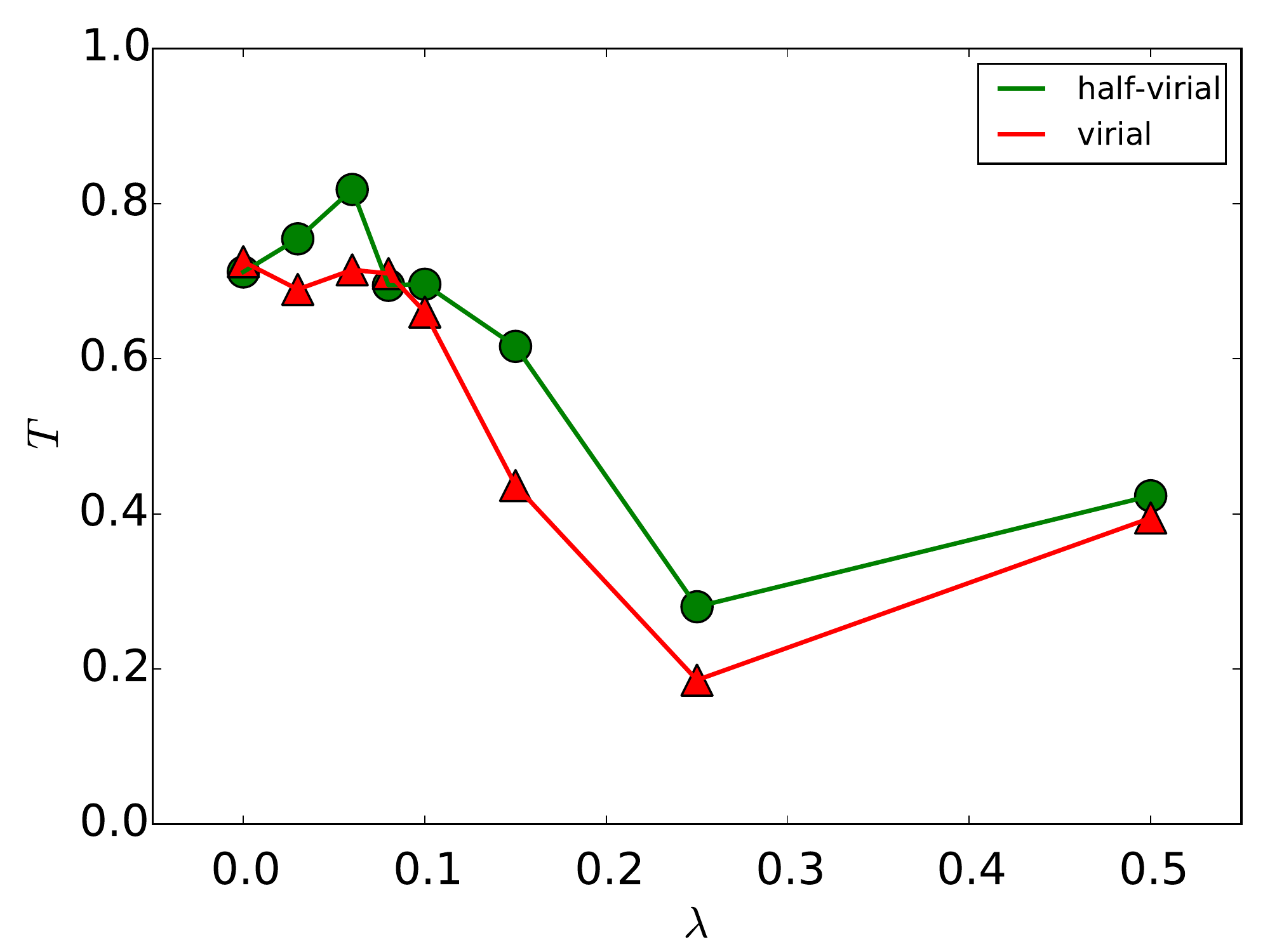}
\caption{Triaxality versus initial spin values for matter within a virial radius and half a virial radius. Note that $T$ values close to 1 are oblate, about 0.5 are triaxial, and close to 0 are prolate. The virialized components appear to shift from a triaxial or partially prolate to oblate shaped over the spin range {$[0.00,0.25]$}, with triaxality being resumed for $\lambda=0.50$. This departure from the trend may due to departure from the small angular momentum energy limit leading to increased extent in-plane.  }
\label{fig:fig5}
\end{center}
\end{figure}

\begin{figure}[]
\begin{center}
\includegraphics[width=8.5cm]{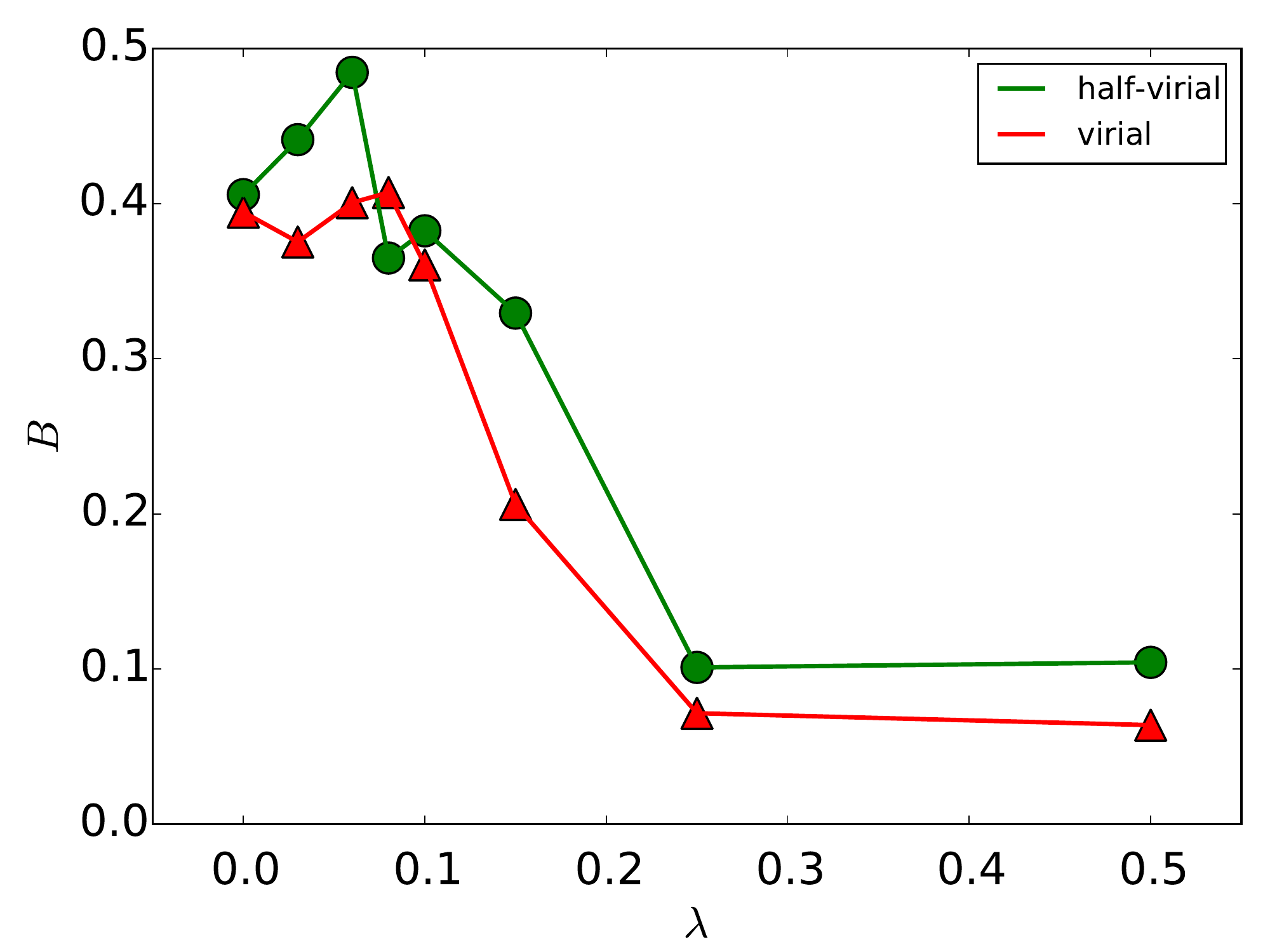}
\caption{Bar parameter versus spin values for particle components within a virial radius and half a virial radius. The bar weakens with increasing spin to $\lambda=0.25$.    }
\label{fig:fig6}
\end{center}
\end{figure}

The difference between the profiles \eqref{Bul_prof} and \eqref{j_prof} points at the existence of a retrograde component to the halo, Fig.~\ref{fig:fig3}, which was not present in the initial configuration. This is curious as the approximate cylindrical symmetry of the initial state should translate to a particle-wise near conservation of the $\hat{\mathbf{z}}$ component of angular momentum. Instead, these halos form a strong bar structure in the rotation plane, Fig.~\ref{fig:fig4}, reminiscent of the radial orbit instability (ROI) \citep{Bellovary, Polyachenko, Barnes1986} in the $\lambda \to 0$ limit. The presence of a strong bar ruins the cylindrical symmetry of the initial conditions, allowing for non-trivial particle-wise angular momentum evolution and the presence of retrograde particles after virialization. The bar is prevalent in all of the generated halos, save for $\lambda=0.25,0.50$ where it weakens significantly. The strength of the bar can, in part, be quantified in terms of its triaxial factor \citep{BT2008}
\begin{equation}
T = \frac{\gamma_1-\gamma_2}{\gamma_1-\gamma_3}, 
\end{equation}
where $\{\gamma_i\}$ are the eigenvalues of the halo's mass quadrupole ordered from largest to smallest, Fig.~\ref{fig:fig5}. Note that the lack of constraint on the principal axes adds ambiguity to whether or not this figure measures the strength of axial symmetry breaking. Since we are only interested in elongation in the plane normal to the z-axis, a more relevant measure of bar strength is in the bar parameter \citep{BT2008}
\begin{equation}
B = \frac{\chi_1- \chi_2}{\chi_1+ \chi_2},
\end{equation}
where $\{ \chi_1, \chi_2 \}$ are the eigenvalues of the projected trace-full quadrupole moment sorted in decreasing value, Fig.~\ref{fig:fig6}.

\begin{figure}[]
\begin{center}
\includegraphics[width=8.5cm]{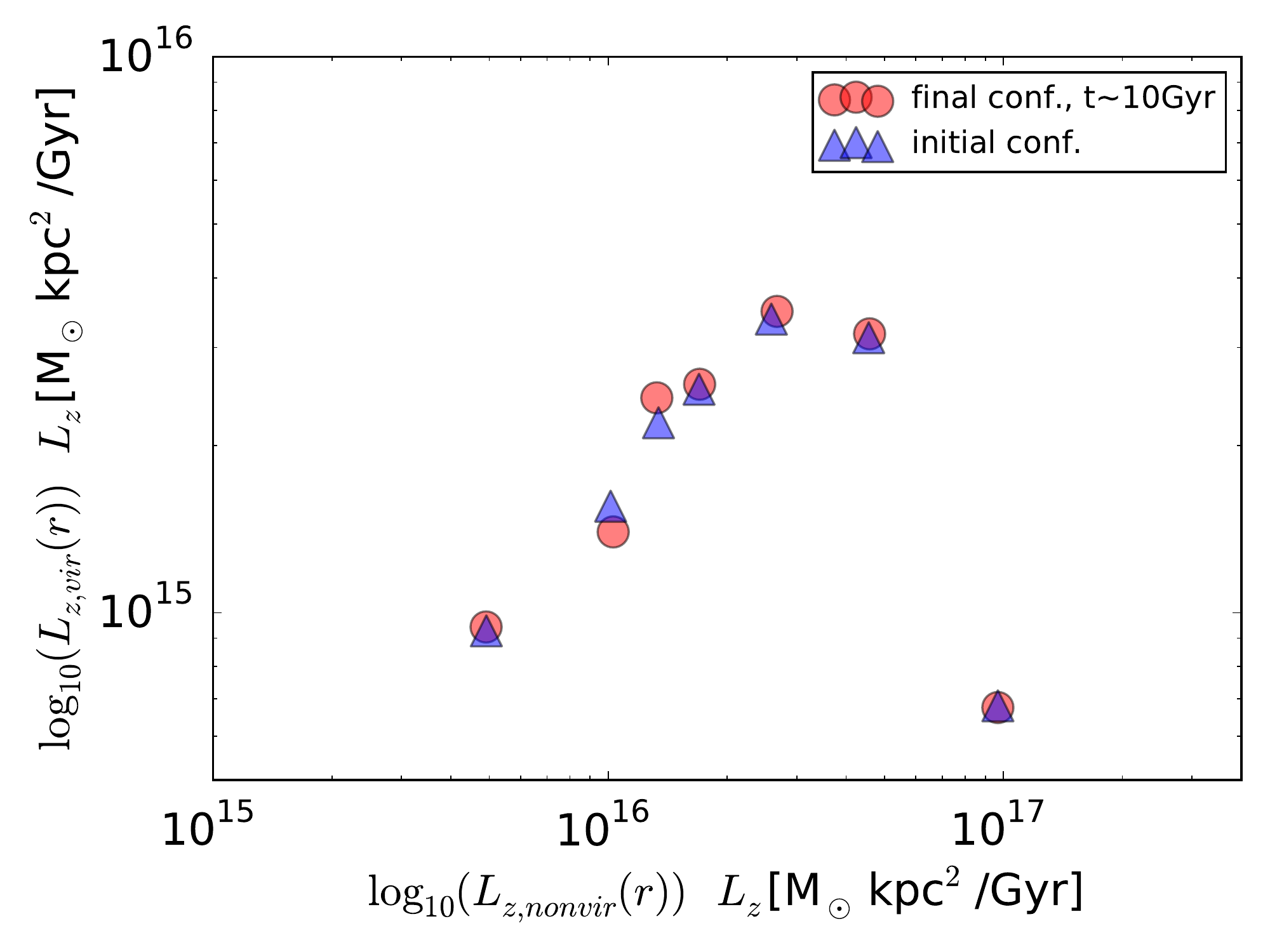}
\caption{Total virialized specific angular momentum versus total nonvirialized specific angular momentum for the initial and final configurations of each spin. The near overlap of the initial and final configurations indicate that angular momentum is almost exactly conserved within the virialized and nonvirialized components \textit{individually}; little to no transfer of angular momentum occurred between the components during the collapse. {The spin values of the points increase from left to right. Note that the spin-less simulation's angular momenta are far smaller than the other runs, and therefore do not appear.} }
\label{fig:fig7}   
\end{center}
\end{figure}

Despite the breaking of cylindrical symmetry, the transfer of angular momentum between virialized and non-virialized components appears to be minimal{ over the initial and final configurations}, Fig.~\ref{fig:fig7}. This effective lack of coupling implies that the instability can exist in isolation, similar to the ROI. We can also see that the amount of retrograde material in the halos is significant, and it can be shown that the density profiles for prograde and retrograde components look very similar for barred halos, and only diverge when the bar weakens.

There appears to be a theme here: for our class of initial distributions, the universality of the density and angular momentum profile shapes persist if and only if a strong bar is present.

\section{Discussion}
\label{Discussion}

The NFW and enclosed mass profiles appear to be robust over a wide range of spin values (Fig.~\ref{fig:fig1}, \ref{fig:fig2}). Also present over these spins is a strong in-plane bar, which contains large amounts of prograde and retrograde material (Fig.~\ref{fig:fig3}). The formation of the bar is analogous to the structure formed by an ROI where a slight breaking of a distribution's spherical or cylindrical symmetry evolves exponentially into a significant asymmetry. \citet{Bellovary} (BDB) showed that for a non-rotating spherical mass, velocity perturbations of the 3D thermal variety disrupted the instability, but purely radial dispersions were ineffective. BDB also made the claim that the ROI is a key physical mechanism contributing to the nearly universal profiles of simulated DM halos.

\citet{BT2008} also parameterize the instability in the non-rotating small angular momentum limit: by relating a spherical distribution's tangential velocity dispersion to spherical stability
\begin{equation}
\vartheta^2 \ge L^2/(G \rho r^2 I) 
\end{equation}
where $L$ is the typical angular momentum per orbit, $\rho$ is the typical density, $r$ is the radial extent of the halo, and $I=T_r/(\partial_L \Delta \phi)|_{L=0}$. To form a distinct bar, $\vartheta$ must be less than $1$. Other studies of this instability \citep{Polyachenko, Barnes1986} are performed in a similar limit. Our work appears to expand triaxial halo formation conditions to include those with coherent tangential components.

The apparent correlation of bar strength with the consistency of the universal profiles is an important step towards understanding the profiles' limits. Equally important is the point that the geometry forms in isolation, with a lack of angular momentum exchange between the virialized and nonvirialized regions. A commonality between the ROI and the bars formed in this study is the high angular momentum dispersion relative to the averaged values, which tapers off as the spin is increased. The dispersions are also seen to decrease in size as the bars weaken (Fig.~\ref{fig:fig5}, \ref{fig:fig6}). If we are to make the connection between the ROI and the instabilities that we form, this observed trend is in violation of the conclusions of Binney and Tremaine that stability against the ROI is driven by tangential velocity dispersion. As their conclusions were derived under different conditions, namely in a non-rotating small angular momentum limit, tangential dispersion alone may be insufficient to determine a halo's stability.

Cylindrical symmetry in the isolated halo may be restored to some degree by the presence of a strong central mass such as a BH or heavy baryonic bulge. Such objects do not exist for certain classes of dwarf galaxies, making our study particularly pertinent to them. Such dwarfs may still exhibit DM cores, which must overcome this propensity for bar creation.

\section{Summary}
\label{Summary}

In this letter we present a study of structure formation on isolated spherical distributions over a range of solid-body spins. This is accomplished via ChaNGa's gravity solvers provided with ICInG N-body {initial} conditions. Our analysis of the collapses show that, for reasonable spin values, the radial density and angular momentum enclosed mass profiles are consolidated and closely conform to NFW and BDK-like profiles respectively. Further investigation into this conspiracy reveals that while there is minimal exchange of angular momentum between the virialized halo and the non-virialized remainder, a great dispersion of angular momentum within the halo occurs, which includes retrograde material. As no retrograde material exists at the onset of the collapses, a breakdown of the axial symmetry must occur. The breakdown takes the form of a bar-like structure, which correlates strongly to the persistence of the universal $\rho(r)$ and $M_{\text{enc}}(j)$ profiles.

Based on these findings, we speculate that the NFW and BDK profiles are robust over the range of probable halo spins due to susceptibility to an instability that bears resemblance to the ROI. The breakdown of such an instability is consistent with the departure from a low angular momentum limit. A study of the robustness of axial symmetry and spin provide insight into the nature of this instability and its profile. Also, for galaxies with a strong central baryonic mass, a study of DM-only collapse with enforced cylindrical symmetry is needed to eliminate spin as a possible solution to the core-cusp problem.

Ultimately, in the larger picture of halo formation, where a cosmological setting leads to hierarchical halo formation, the role of these instabilities becomes less clear. A halo's history consists of periods of gentle minor mergers, interrupted occasionally by more violent major mergers. \citet{Bellovary} performs preliminary simulations of halos subject to controlled minor and major mergers to investigate the realistic role of the ROI, where they speculate that the ROI does not play a significant role in the subsequent relaxation process after a major merger. However, they also state that there are indications of an operational ROI during the minor merger periods, acting on nearly radial tidal streams associated with disrupted subhalos. A similar study with non-radial inflow and torquing may provide insight into the rotational regime.

\section{Acknowledgements}
\label{Acknowledgements}

We gratefully acknowledge the support of the U.S. Department of Energy office of High Energy Physics and the National Science Foundation. TQ was supported in part by the NSF grant AST-1311956. EL and LR were supported by the DOE grant DE-SC0011665.

\end{document}